\documentclass{elsart}

\usepackage{graphicx, amssymb, subfigure}

\begin{document}

\begin{frontmatter}

\title{Muon Charge Information from Geomagnetic Deviation in Inclined Extensive Air Showers}

\author[PKU]{BingKan Xue}
\author[CCAST,PKU]{Bo-Qiang Ma \thanksref{ma}}
\thanks[ma]{Corresponding author. E-mail address: mabq@phy.pku.edu.cn}
\address[PKU]{Department of Physics, Peking University, Beijing 100871,
China} \address[CCAST]{CCAST (World Laboratory), P.O.~Box 8730,
Beijing 100080, China}

\begin{abstract}
We propose to extract the charge information of high energy muons in
very inclined extensive air showers by analyzing their relative
lateral positions in the shower transverse plane. We calculate the
muon lateral deviation under the geomagnetic field and compare it to
dispersive deviations from other causes. By our criterion of
resolvability, positive and negative muons with energies above
$10^4$\,GeV will be clearly separated into two lobes if the shower
zenith angle is larger than $70^\circ$. Thus we suggest a possible
approach to measure the $\mu^+ / \mu^-$ ratio for high energy muons.
\end{abstract}


\end{frontmatter}

\section{Introduction}

The study of cosmic rays with primary energies above $10^5$\,GeV are
typically based on the measurements of extensive air showers (EAS)
that they initiate in the atmosphere. The ground detector array
records the secondary particles produced in shower cascades,
including photons, electrons (positrons), muons, and some hadrons.
Then their arrival times and density profiles are used to infer the
primary energy and composition of the incident cosmic ray particle
\cite{Anchordoqui:2004xb}, usually through comparison with simulated
results. EAS events initiated by primary particles with energies
above $10^{10.9}$\,GeV, the so called GZK cutoff energy
\cite{Greisen:1966jv}, have already been reported
\cite{Bird:1994uy}. Questions about the composition of such ultra
high energy (UHE) cosmic ray particles are still open to
investigations \cite{Stanev:2004kk}.

Photons, electrons and positrons are the most numerous secondary
particles in an EAS event. However, for very inclined showers which
will be concerned in this article, these electromagnetic components
would travel a long slant distance and are almost completely
absorbed before they reach the ground \cite{Zas:2005zz}. On the
other hand, muons are decay products of charged mesons in shower
hadronic cascades. Most high energy muons survive their propagation
through the slant atmospheric depth, during which they lose
typically a few tens of GeV's energy \cite{Eidelman:2004wy}. These
high energy muons carry important information about the nature of
the primary cosmic ray hadron, which will be extracted from their
energy spectrum and lateral distribution.

The ratio of positive versus negative muons $\mu^+ / \mu^-$ is a
significant quantity which can help to discern the primary
composition \cite{Adair:1977se}, and at high energies this charge
ratio also reflects important features of hadronic meson production
in cosmic ray collisions \cite{Vulpescu:1998hm}. In order to obtain
such muon charge information, we would need a way to distinguish
between positive and negative high energy muons. Unfortunately,
existing muon detectors available at shower arrays, usually
scintillators and water \v Cerenkov detectors
\cite{Anchordoqui:2004xb}, are not commonly equipped with magnetized
steel to differentiate the muon charges. Even if they were, the
limited region of the magnetic field prevents definite determination
of high energy muons' track curvature.

This invites us to think of the geomagnetic field as a huge natural
detector for muon charge information. Apparently, after being
produced high in the atmosphere, a positively charged muon would
bend east on its way down while a negatively charged muon would bend
west, introducing an asymmetry into the density profile of the
shower front. If their separation is large enough as compared with
other circularly symmetric ``background'' deviations, it will be
possible to distinguish the positive muons from the negative ones.

To see such an effect, we may need a detailed simulation of inclined
air showers, which keeps track of both the charges and lateral
positions of numerous muons produced in the shower cascades.
However, a simple model will be enough to analyze the practicability
of our approach without cumbersome computations. We will calculate
the muon lateral distribution and see its dependence on the shower
zenith angle. By introducing a quantitative criterion to the muon
energy, we can get an estimation of the condition for an unambiguous
geomagnetic effect.

The article is organized as follows. We start by a general
consideration of muon production and lateral deviations in
section~\ref{sec:gen}. The expressions for muon lateral deviations
are derived explicitly in section~\ref{sec:dev}. Then we present a
revised Heitler model in section~\ref{sec:mod} to complete the
calculation of muon lateral distribution. Our criterion is applied
and some typical situations are discussed. Finally in
section~\ref{sec:num} we calculate the muon energy spectrum based on
our model, and obtain a rough image of the muon number density. We
will conclude with our proposed approach to measure the muon charge
ratio in section~\ref{sec:sum}.

\section{General Features}
\label{sec:gen}

To understand the development of an extensive air shower and the
major processes in which muons are involved, we first introduce a
Heitler model \cite{Matthews:2005sd}, which describes the air shower
in a simple and analytical way but captures the main features of
both electromagnetic cascades and hadronic multiparticle
productions. Since our interest concentrates on muon production, we
will only review the hadronic part briefly.

Let us assume that charged hadrons undergo multiparticle productions
once they travel an interaction length $\lambda_I$, producing a new
generation of  $N_{ch}$ charged pions and $\frac{1}{2}N_{ch}$
neutral pions with equal energies, where $N_{ch}$ is the
multiplicity of charged particles in the hadron-air interaction. The
neutral pions decay immediately to photons, initiating
electromagnetic subshowers, while the charged pions traverse another
atmospheric depth $\lambda_I$ and experience multiparticle
productions of their own. For interactions in the energy range
$10-1000$\,GeV, both $\lambda_I$ and $N_{ch}$ can be approximated as
constants \cite{Matthews:2005sd}, where $\lambda_I \approx
120$\,g/cm$^2$, and $N_{ch} \approx 10$.

Consider a single cosmic ray nucleon with primary energy $E_0$
incident into the atmosphere. After $n$ interactions, there are $N_n
= \left( N_{ch}\right)^n$ charged pions in total with equal energies
\begin{equation}
E_n = E_0/ \Big( \frac{3}{2} N_{ch} \Big)^n,\label{eq:EnH}
\end{equation}
and the shower front has traversed an atmospheric depth
\begin{equation}
X_n = n \lambda_I.\label{eq:XnH}
\end{equation}
Let us solve Eq.~(\ref{eq:EnH}) for $n$ and replace it in
Eq.~(\ref{eq:XnH}), so that we have a continuous function of $X$ on
$E$,
\begin{equation}
X(E) = \lambda_I \frac{\ln(E_0/E)}{\ln(\frac{3}{2}N_{ch})}.
\label{eq:XEH}
\end{equation}
If we adopt the isothermal atmospheric density $\rho(h) = \rho_0
e^{-h/h_0}$ with $\rho_0 = 1.225 \times 10^{-3}$\,g/cm$^3$ and $h_0
= 8.4$\,km \cite{Anchordoqui:2004xb}, then the atmospheric depth $X$
can be related to height $H$ (assuming a vertical shower) by
\begin{equation}
X(H) = \int_H^\infty \rho(h)dh = X_0 e^{-H/h_0}, \label{eq:XH}
\end{equation}
where the total atmospheric depth $X_0 = \rho_0 h_0 =
1030$\,g/cm$^2$. Eqs. (\ref{eq:XEH}) and (\ref{eq:XH}) can be solved
to give a relation between the pion energy $E$ and the height $H$,
\begin{equation}
H(E) = h_0 \ln \left( \frac{X_0}{\lambda_I}
\frac{\ln(\frac{3}{2}N_{ch})}{\ln(E_0/E)} \right). \label{eq:HEH}
\end{equation}

We are now in a position to consider muon production and propagation
in the air shower. Although in the above Heitler model we assumed
that all charged pions experience hadronic interactions after they
travel an atmospheric depth $\lambda_I$, they actually have certain
probability of decaying to muons before they interact. When that
happens, the muon would inherit about $1/1.27$ of the pion energy on
average \cite{Penchev:1999tv}. Simply inserting a factor $1.27$ in
front of $E$ in Eq.(\ref{eq:HEH}), we find the relation between the
production height and the muon energy. It is apparent that more
energetic muons are produced higher in the atmosphere and travel a
longer distance before they reach the ground.

We should expect that the decay probability of pions in the first
generations is extremely small, with few highest energy muons
actually produced. The probability approaches $1$ as the pion energy
rapidly drops, so we set a critical energy $\xi^\pi_c \sim 10$\,GeV
below which all pions are supposed to decay rather than interact
\cite{Matthews:2005sd}, and the hadronic multiplication is cut off.

To see the muon lateral distribution, we make the approximation that
their lateral deviations from different causes can be calculated
independently. Let us consider muons produced at a same height with
equal energies. If not for the deviational effects, they would all
have landed at a single point, i.e. the intersection of shower axis
and the ground plane. However, due to their transverse momenta from
production and multiple Coulomb scattering with air nuclei
\cite{Knapp:2002vs}, these muons would spread isotropically in a
plane perpendicular to the shower axis as they travel down,
resulting in a circular distribution as a background.

In the presence of the geomagnetic field, positive and negative
muons would bend in opposite directions perpendicular to the shower
axis. This extra geomagnetic deviation splits the original landing
point into two separate centers, one for the beam of positive muons
and the other negative. Both beams experience lateral dispersions
due to their transverse momenta and multiple scattering, which
superposes the background circular distribution onto each center,
resulting in a distinct double-lobed feature.

Conceivably, if the separation between the two centers is too small,
then we can hardly resolve one lobe from the other. On the contrary,
if we have a separation much larger than the radial extent of either
circular distribution, hence little overlap between the two lobes,
then we can be confident that each lobe mainly consists of muons
with the same charge (positive or negative). In analogy to
Rayleigh's criterion in optics, we define the condition for
resolvability to be that the separation of the two centers
\emph{exceeds twice the attenuation radius} of each background
distribution.

In order that the separation be large, we will seek a big
geomagnetic deviation, which implies a long muon trajectory. In this
case inclined air showers are more favorable than vertical showers,
because the total distance that a muon travels is magnified by a
factor of $\sec\theta$, where $\theta$ is the zenith angle of the
shower axis. Our above analysis can be applied directly to inclined
showers if we restrict our conclusions in the plane perpendicular to
the shower axis (the \emph{transverse plane}) instead of the ground
plane. A method for projections between these two planes can be
found in \cite{Ave:2000xs}.

\section{Muon Lateral Deviations}
\label{sec:dev}

Let us consider muons produced at height $H$ with energy $E$ in an
inclined air shower with zenith angle $\theta$. The total length $L$
(\emph{slant distance}) of muon trajectory along the shower axis is
therefore $H\sec\theta$. For high energy muons produced early in the
cascade, the lateral distance from their production sites to the
shower axis can be safely neglected, so we think of these muons as
being produced on the shower axis and then starting to deviate from
it.

We shall calculate the muon lateral deviations in the transverse
plane due to their transverse momenta, multiple scattering and
geomagnetic bending separately. Since we are interested in muons
with sufficiently high energies, their energy losses during
propagation are relatively small and will be neglected.

\subsection{Transverse momentum}
The charged pions from hadronic multiparticle production have a
relatively broad distribution in their transverse momenta. When they
decay to muons, the transverse momentum distribution is largely
maintained. For convenience in calculation, we can approximate this
$p_\textrm{\tiny T}$ distribution by a Gaussian form
\cite{Aly:1964},
\begin{equation}
f(p_\textrm{\tiny T})dp_\textrm{\tiny T} = \frac{2\,p_\textrm{\tiny
T}}{\langle p_\textrm{\tiny T}^2 \rangle} ~e^{-\frac{p_\textrm{\tiny
T}^2}{\langle p_\textrm{\tiny T}^2 \rangle}} dp_\textrm{\tiny T}.
\label{eq:fp}
\end{equation}

If we denote the lateral deviation due to $p_\textrm{\tiny T}$ by
$r_p$, then we have $\frac{r_p}{L} = \frac{p_\textrm{\tiny T}}{p}$,
where $p \simeq E/c$ is the longitudinal momentum of the muon.
Replacing $p_\textrm{\tiny T}$ with $\frac{E}{L} ~r_p$ in
Eq.~(\ref{eq:fp}), we find the radial distribution of muons in the
transverse plane,
\begin{equation}
f(r_p)dr_p = \frac{E^2}{L^2} \frac{2\,r_p}{\langle p_\textrm{\tiny
T}^2 \rangle} ~e^{-\frac{E^2}{L^2} \frac{r_p^2}{\langle
p_\textrm{\tiny T}^2 \rangle}} dr_p, \label{eq:frp}
\end{equation}
or in $(x,y)$ coordinates,
\begin{equation}
f(x,y)dxdy = \frac{1}{2\pi\sigma_p^2}
~e^{-\frac{x^2+y^2}{2\sigma_p^2}} dxdy, \label{eq:fyp}
\end{equation}
whose standard deviation is
\begin{equation}
\sigma_p = \frac{L}{E} \,\frac{\langle p_\textrm{\tiny T}^2 \rangle
^{1/2}}{\sqrt{2}}. \label{eq:sigmap}
\end{equation}

Simulation of hadronic multiparticle production with
QGSJet-\uppercase\expandafter{\romannumeral2} shows that in the
energy range $10^4-10^{10}$\,GeV $\langle p_\textrm{\tiny T}
\rangle$ is almost a constant $\approx 0.5$\,GeV
\cite{Horandel:2003vu}, so that from Eq.~(\ref{eq:fp}),
$\frac{\langle p_\textrm{\tiny T}^2 \rangle ^{1/2}}{\sqrt{2}} =
\sqrt{\frac{2}{\pi}} \langle p_\textrm{\tiny T} \rangle \approx
0.4$\,GeV. Adopting this value in Eq.~(\ref{eq:sigmap}) gives
\begin{equation}
\sigma_p = \frac{400\,(L/\textrm{km})}{(E/\textrm{GeV})}
~\textrm{m}. \label{eq:sigmap1}
\end{equation}

\subsection{Multiple scattering}

Under the approximation that different lateral deviations can be
considered independently, we will neglect the muons' initial
transverse momenta when calculating their deviations caused by
multiple scattering, assuming the muons to be moving along the
shower axis when they are produced.

After traversing an atmospheric depth $\Delta X$, some muons will be
deflected by an angle $\phi$ as a result of multiple Coulomb
scattering with air nuclei. The deflection angle $\phi$ has a
distribution that is nearly Gaussian \cite{Ayre:1972nx},
\begin{equation}
f(\phi)d\phi = \frac{2\phi}{\langle \phi^2 \rangle}
~e^{-\frac{\phi^2}{\langle \phi^2 \rangle}} d\phi. \label{eq:ftheta}
\end{equation}
The mean squared deflection angle is related to the traversed
atmospheric depth by
\begin{equation}
\langle \phi^2 \rangle = \left( \frac{E_s}{p \beta c} \right)^2
\frac{\Delta X}{x_0}, \label{eq:phi}
\end{equation}
where $E_s = 21$\,MeV, and the radiation length $x_0 =
36.7$\,g/cm$^2$ in air. For muons with energies $\gtrsim$ TeV, we
have $\beta \simeq 1$ and $pc \simeq E$, so that
\begin{equation}
\langle \phi^2 \rangle = \left( \frac{0.021}{E/\textrm{GeV}} \right)
^2 \,\frac{(\Delta X / \textrm{g\,cm}^{-2})}{36.7}. \label{eq:phi1}
\end{equation}

We can use Eq.~(\ref{eq:phi1}) and $X = X_0
\,e^{-\frac{l\cos\theta}{h_0}}$ from isothermal atmosphere to
calculate the muon lateral deviation $r_s$. The result is also a
Gaussian distribution like Eq.~(\ref{eq:fyp}), but with standard
deviation
\begin{equation}
\sigma_s = \frac{\langle r_s^2 \rangle ^{1/2}}{\sqrt{2}} =
\frac{0.021}{\sqrt{2}E} \sqrt{\frac{\rho_0}{x_0}} \left[ \left(
\frac{h_0}{\cos\theta} \right) L^2 - \left( \frac{h_0}{\cos\theta}
\right)^2 2L + 2\left( \frac{h_0}{\cos\theta} \right)^3 \left(
1-e^{-\frac{L\cos\theta}{h_0}} \right) \right]^{1/2}.
\label{eq:sigmas}
\end{equation}

Now the total background dispersion is the superposition of the
above two kinds of lateral deviations. Let us ``add together'' these
two Gaussian distributions, which results in another Gaussian
distribution with a combined standard deviation,
\begin{equation}
f(x,y) = \frac{1}{2\pi\sigma^2} ~e^{-\frac{x^2+y^2}{2\sigma^2}},
\quad \sigma = \sqrt{\sigma_p^2 + \sigma_s^2}. \label{eq:f0}
\end{equation}

\subsection{Geomagnetic deviation}

In this calculation we leave aside the effects of both transverse
momenta and multiple scattering. Let us decompose the geomagnetic
field $\vec{B}$ into components $\vec{B_\parallel}$ parallel to the
shower axis and $\vec{B_\perp}$ perpendicular to it, then we define
coordinates $(x,y)$ in the transverse plane such that $y$ axis is in
the direction of $\vec{B_\perp}$. Since an EAS event takes place
within a small region of Earth's surface, the magnetic field
$\vec{B}$ is almost a constant for a specific shower location.
Moreover, for very inclined air showers $B_\perp$ is approximately
independent of azimuth \cite{Ave:2000xs}.

Therefore, we will simply set $B_\perp$ to a reasonable fixed value
$40\,\mu$T, and the muon trajectory can be well approximated by an
arc in the plane containing the $x$ and the shower axis. The radius
of curvature is
\begin{equation}
R = \frac{p}{eB_\perp} = \frac{E/c}{eB_\perp} =
\frac{(E/\textrm{GeV})}{3\,(B_\perp/\mu \textrm{T})} ~10^4
\,\textrm{km}, \label{eq:R}
\end{equation}
hence the lateral deviation
\begin{equation}
x_g = R \left[ 1 - \Big( 1 - \frac{L^2}{R^2} \Big) ^{1/2} \right]
\approx \frac{L^2}{2R} =
\frac{3\,(B_\perp/\mu\textrm{T})\,(L/\textrm{km})^2}{20\,(E/\textrm{GeV})}
~\textrm{m}, \label{eq:xg}
\end{equation}
where we have expanded the bracket to first order in $\left(
\frac{L}{R} \right) ^2$, which is small as seen from
Eq.~(\ref{eq:R}). Since positive and negative muons deviate in
opposite directions, their separation is simply twice $x_g$. Putting
$B_\perp = 40 \,\mu$T gives
\begin{equation}
s = \frac{12\,(L/\textrm{km})^2}{(E/\textrm{GeV})} ~\textrm{m}.
\label{eq:s}
\end{equation}

To see the muon lateral distribution, we now superpose the
background deviation Eq.~(\ref{eq:f0}) onto two separated centers
for the opposite muon charges. Let us denote the muon charge ratio
$\mu^+ / \mu^-$ by $R_\mu$, so that $\frac{R_\mu}{1+R_\mu}$ of the
total muons are positive and the rest $\frac{1}{1+R_\mu}$ are
negative. Putting the lobe of positive muons on the right and the
negative on the left, we finally arrive at the distribution
\begin{equation}
f(x,y) = \frac{1}{2\pi\sigma^2} \left( \frac{1}{1+R_\mu}
~e^{-\frac{(x+\frac{s}{2})^2+y^2}{2\sigma^2}} +
\frac{R_\mu}{1+R_\mu} ~e^{-\frac{(x-\frac{s}{2})^2+y^2}{2\sigma^2}}
\right), \label{eq:fR}
\end{equation}
where $\sigma = \sqrt{\sigma_s^2 + \sigma_p^2}$ is given by Eqs.
(\ref{eq:sigmap}) \& (\ref{eq:sigmas}), and $s$ by Eq.~(\ref{eq:s}).

Note that $R_\mu$ is found to be almost a constant around $1.268$ in
the energy range $10-300$\,GeV \cite{Hebbeker:2001dn}, yet there is
no systematic measurement of its value above that energy. Since
small changes in the value of $R_\mu$ do not affect the validity of
our approach, here we first set $R_\mu$ to $1$ for symmetry and
simplicity, so that
\begin{equation}
f(x,y) = \frac{1}{4\pi\sigma^2} \left(
e^{-\frac{(x+\frac{s}{2})^2+y^2}{2\sigma^2}} +
e^{-\frac{(x-\frac{s}{2})^2+y^2}{2\sigma^2}} \right). \label{eq:f}
\end{equation}
We will propose an approach to measure the precise value of $R_\mu$
later.

Eq.~(\ref{eq:f}) is exactly the double-lobed distribution we
expected. If we take $\sigma$ to be the attenuation radius of either
lobe, then our criterion of muon charge resolvability is
$s>2\sigma$. Since longer muon trajectories correspond to higher
production sites and greater muon energies, the slant distance $L$
in Eqs.~(\ref{eq:sigmap}), (\ref{eq:sigmas}) and (\ref{eq:s}) is a
function of $E$ implicitly. Therefore, to see the muon energy that
meets our criterion, we will need an appropriate $L$\,-\,$E$
relation to convert the dependence of $s$ and $\sigma$ on the slant
distance to that on muon energy.

\section{Revised Heitler Model}
\label{sec:mod}

The advantage of the Heitler model briefly discussed in section 2 is
that it plainly demonstrates the development of extensive air
showers with simplest calculations. However, when we consider air
showers with energies several orders beyond TeV, some basic
assumptions of the original Heitler model no longer hold strictly.
In fact, as the hadron energy rises from a few GeV up to over
$10^{10}$\,GeV, the multiplicity of hadron-air interactions
increases rapidly while the interaction length of the hadrons
decreases by more than a half \cite{Ostapchenko:2004qz}.
Consequently, some revisions are necessary to extend the Heitler
model to higher energies, which will result in a more realistic
$L$\,-\,$E$ relation than Eq.~(\ref{eq:HEH}).

\begin{figure}
\begin{center}
\includegraphics{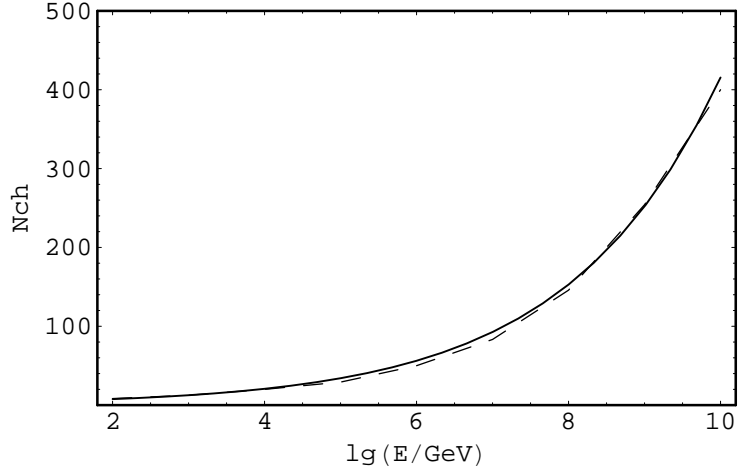}
\caption{Multiplicity of charged particles in pion-air interactions
from QGSJet-\uppercase\expandafter{\romannumeral2} (dashed curve)
and the empirical formula (\ref{eq:Nch}) with $A=2.8,\ \alpha=0.5$
(solid curve).} \label{fig:Nch}
\end{center}
\end{figure}

Fig.~\ref{fig:Nch} shows the multiplicity of charged particles
$N_{ch}$ in pion-air interactions predicted by simulations using the
QGSJet-\uppercase\expandafter{\romannumeral2} model (dashed curve).
We can see an exponential growth of $N_{ch}$ with the logarithm of
pion energy. This trend can be fitted by a function like
\begin{equation}
N_{ch}(E) = A \,e\,^{\alpha \lg E}, \label{eq:Nch}
\end{equation}
with $E$ always expressed in GeV. A best fit gives the parameters $A
= 2.8$ and $\alpha = 0.5$ (solid curve in Fig.~\ref{fig:Nch}), which
compares well with the simulated result.

\begin{figure}
\begin{center}
\includegraphics{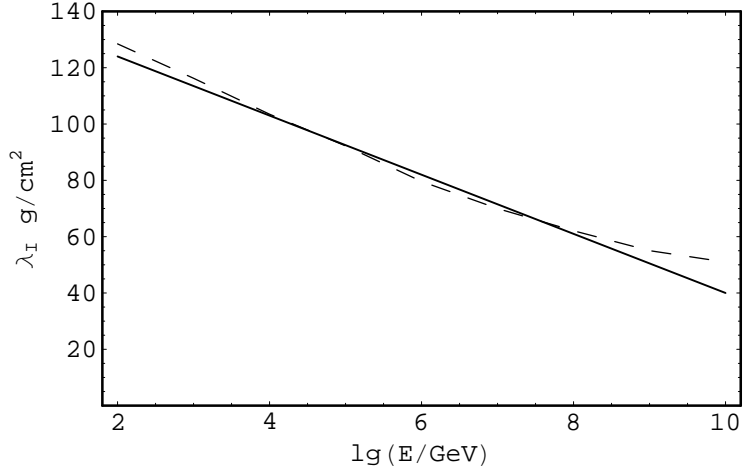}
\caption{Pion-air interaction length from
QGSJet-\uppercase\expandafter{\romannumeral2} (dashed curve) and the
empirical formula (\ref{eq:lambda}) with $B=145,\ C=10.5$ (solid
curve).} \label{fig:lam}
\end{center}
\end{figure}

Similarly, we introduce another empirical formula to approximate the
decrease of pion interaction length with energy,
\begin{equation}
\lambda_I(E) = B - C \,\lg E, \label{eq:lambda}
\end{equation}
where $B$ and $C$ are constants chosen to fit the simulated curve.
Fig.~\ref{fig:lam} shows a comparison between our result with best
parameters $B = 145$\,g/cm$^2$, $C = 10.5$\,g/cm$^2$ (solid curve)
and that from QGSJet-\uppercase\expandafter{\romannumeral2} (dashed
curve).

To see the revised Heitler model, we only need to replace $N_{ch}$
and $\lambda_I$ in the original model with the empirical formulas
(\ref{eq:Nch}) and (\ref{eq:lambda}). More specifically, let us
consider the $n$th hadronic interaction. The energy of the charged
pions before this interaction is $E_{n-1} = \big( \frac{2}{3} \big)
^{n-1} \frac{E_0}{N_{n-1}}$, where $E_0$ is the \emph{primary
energy} of the cosmic ray particle, and $N_{n-1}$ is the \emph{total
number of charged pions} before this interaction. By
Eq.~(\ref{eq:Nch}) the \emph{total multiplicity} of this hadronic
interaction is $M_{n-1}=\frac{3}{2} A \,e^{\alpha \lg E_{n-1}}$,
producing $N_n=N_{n-1}\cdot\frac{2}{3}M_{n-1}$ charged pions with
equal energies
\begin{equation}
E_n=\frac{E_{n-1}}{M_{n-1}} = \Big( \frac{2}{3} \Big)^n
~\frac{E_0}{N_n}. \label{eq:En}
\end{equation}
Meanwhile, the parent pion has traversed an atmospheric depth
$\lambda_{n-1}= B - C \,\lg E_{n-1}$ before it interacts, pushing
the shower  front to a greater \emph{total depth} $X_n = X_{n-1} +
\lambda_{n-1}$. As usual, the multiplication of charged pions is
supposed to continue before they reach the critical energy
$\xi^\pi_c \sim 10$\,GeV.

We have to use the above recursive relations repeatedly to derive an
expression of $E_n$ and $X_n$ with regards to $n$ only. Fortunately,
this is possible due to the special forms of empirical formulas
(\ref{eq:Nch}) and (\ref{eq:lambda}). The main steps are:
\begin{eqnarray}
M_n &=& M_0^{\left( 1 - 0.4343\alpha \right)^n} = \Big( \frac{3}{2}A\,e\,
^{\alpha \lg E_0} \Big) ^{(1-0.4343\alpha)^n},\label{eq:Mn}\\
E_n &=& \Big( \frac{3}{2}A \Big) ^{-\frac{1-\left( 1-0.4343\alpha
\right)^n}{0.4343\alpha}} E_0^{\left( 1-0.4343\alpha \right)^n},\label{eq:En1}\\
X_n &=& n \left( \frac{C}{\alpha} \ln \Big( \frac{3}{2}A \Big) + B
\right) + \frac{C}{0.4343\alpha} \left( \lg E_n - \lg E_0
\right).\label{eq:Xn}
\end{eqnarray}
Just like what we did with the original model, we now solve
Eq.~(\ref{eq:En1}) for $n$ and then plug it into Eq.~(\ref{eq:Xn})
to obtain $X$ as a continuous function of $E$. The result is
\begin{equation}
X(E) = \left( \frac{C}{\alpha} \ln \Big( \frac{3}{2}A \Big) + B
\right) \frac{\ln \left( \frac{\frac{1}{\alpha}\ln (\frac{3}{2}A) +
\lg E}{\frac{1}{\alpha}\ln (\frac{3}{2}A) + \lg E_0} \right)}{\ln
\left( 1 - 0.4343\alpha \right)} + \frac{C}{0.4343\alpha} \left( \lg
E - \lg E_0 \right), \label{eq:XE}
\end{equation}
or, adopting the best-fit parameters,
\begin{equation}
X(E) = \left( -21 ~\lg \frac{E_0}{E} + 715.4 ~\ln \frac{2.870 + \lg
E_0}{2.870 + \lg E} \right) ~\textrm{g/cm}^2. \label{eq:XE1}
\end{equation}
Referring to Eq.~(\ref{eq:XH}) for the height $H$, and taking
account of the shower zenith angle $\theta$, we find the slant
distance
\begin{equation}
L = h_0 \sec\theta ~\ln \frac{X_0 \sec\theta}{X(E)},
\end{equation}
where $X(E)$ is given by Eq.~(\ref{eq:XE1}).

\begin{figure}
\begin{center}
\includegraphics{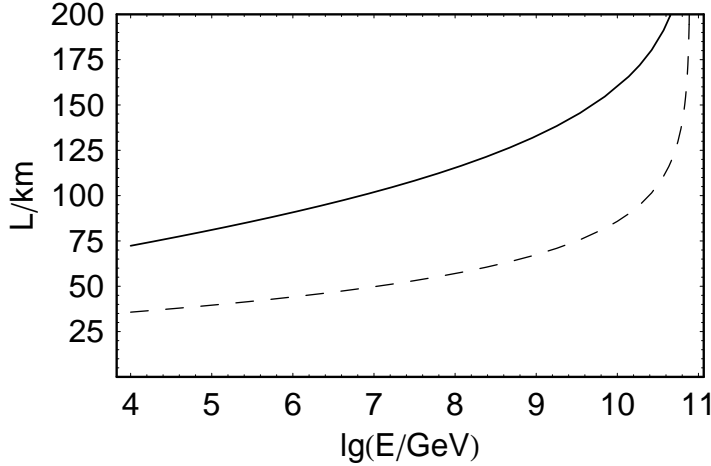}
\caption{Relation between the slant distance of muon production site
and the muon energy for air showers with $E_0=10^{11}$\,GeV and
$\theta=70^\circ$, calculated with the original Heitler model
(dashed curve) and the revised model (solid curve).} \label{fig:LE}
\end{center}
\end{figure}

To make $E$ the energy of the decay muon, we should insert the
factor $1.27$ in this expression, which results in exactly the
$L$\,-\,$E$ relation that we sought in the last section,
\begin{equation}
L(E) = h_0 \sec\theta ~\ln \frac{X_0 \sec\theta}{X(1.27\,E)}.
\label{eq:LE}
\end{equation}
Fig.~\ref{fig:LE} compares the result from the original Heitler
model (dashed curve) with that from the revised one (solid curve),
their difference being so evident that our revision is justified.

Let us set out to use this $L(E)$ function to complete our
calculation of the muon lateral distribution and determine the muon
energy range suitable for our purpose. By replacing $L$ in
expressions (\ref{eq:sigmap}), (\ref{eq:sigmas}) and (\ref{eq:s})
with Eq.~(\ref{eq:LE}), we render the separation $s$ and deviation
$2\sigma$ in Eq.~(\ref{eq:f}) functions of $E$ only. Thus we can
compare the variation of their relative magnitudes with energy.

\begin{figure}
\subfigure[$\theta=70^\circ$]{\label{fig:s70}
\includegraphics[width=0.5\textwidth]{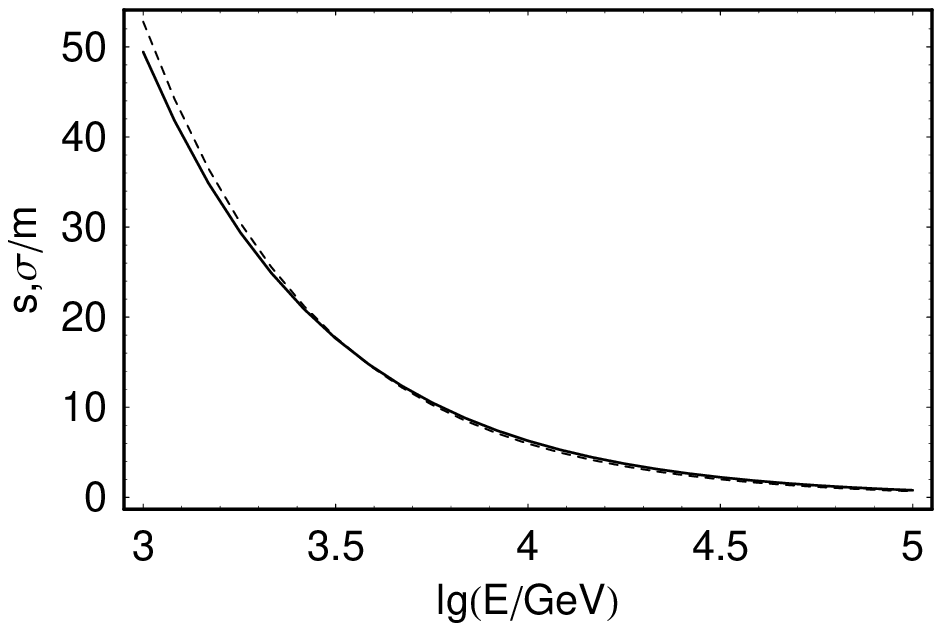}}
\subfigure[$\theta=75^\circ$]{\label{fig:s75}
\includegraphics[width=0.5\textwidth]{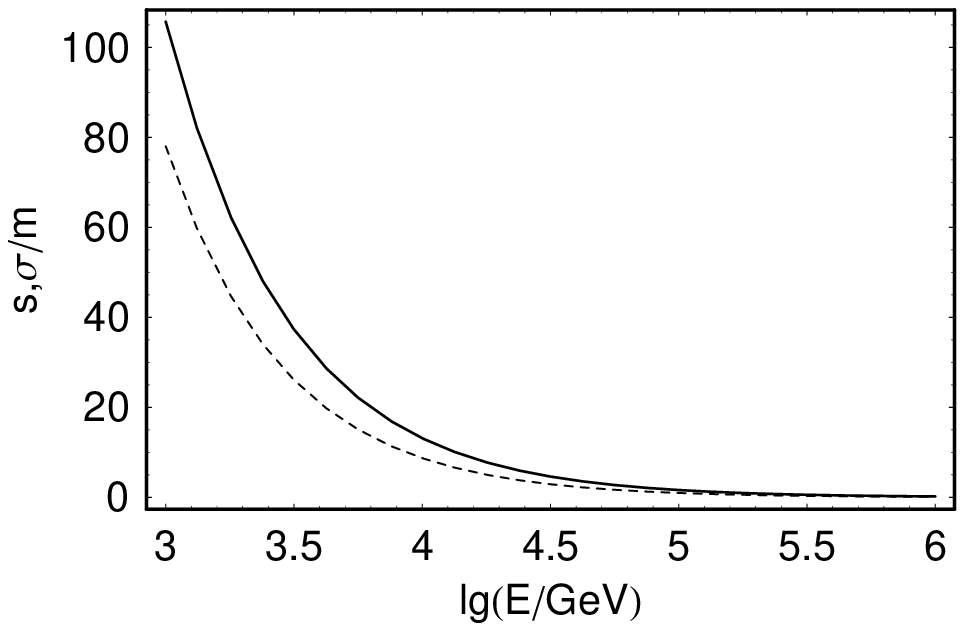}}
\subfigure[$\theta=80^\circ$]{\label{fig:s80}
\includegraphics[width=0.5\textwidth]{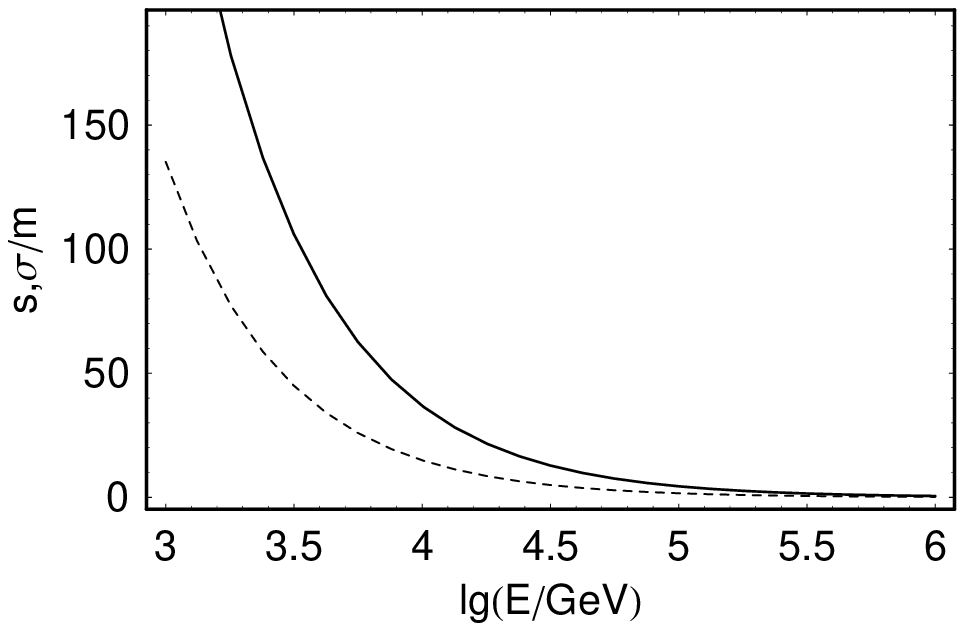}}
\subfigure[$\theta=85^\circ$]{\label{fig:s85}
\includegraphics[width=0.5\textwidth]{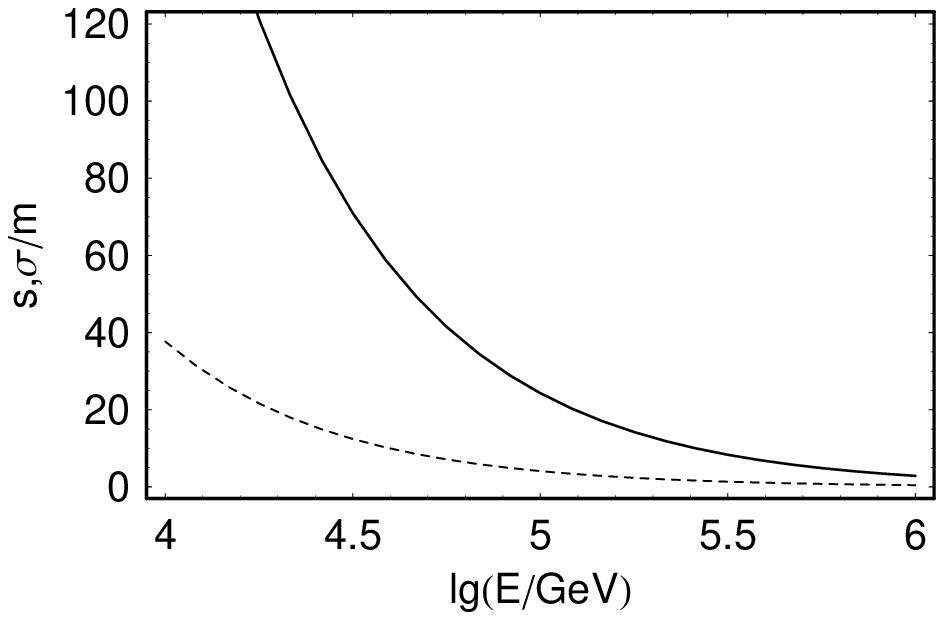}}
\caption{Dependence of the separation $s$ (solid curve) and
attenuation radius $2\sigma$ (dashed curve) in Eq.~(\ref{eq:f}) on
the muon energy, assuming air showers with primary energy
$E_0=10^{11}$\,GeV and at different zenith angles $\theta=
70^\circ,\ 75^\circ,\ 80^\circ,\ 85^\circ$ respectively.}
\end{figure}

\begin{figure}
\subfigure[$\theta=70^\circ$]{\label{fig:f70}
\includegraphics[width=0.5\textwidth]{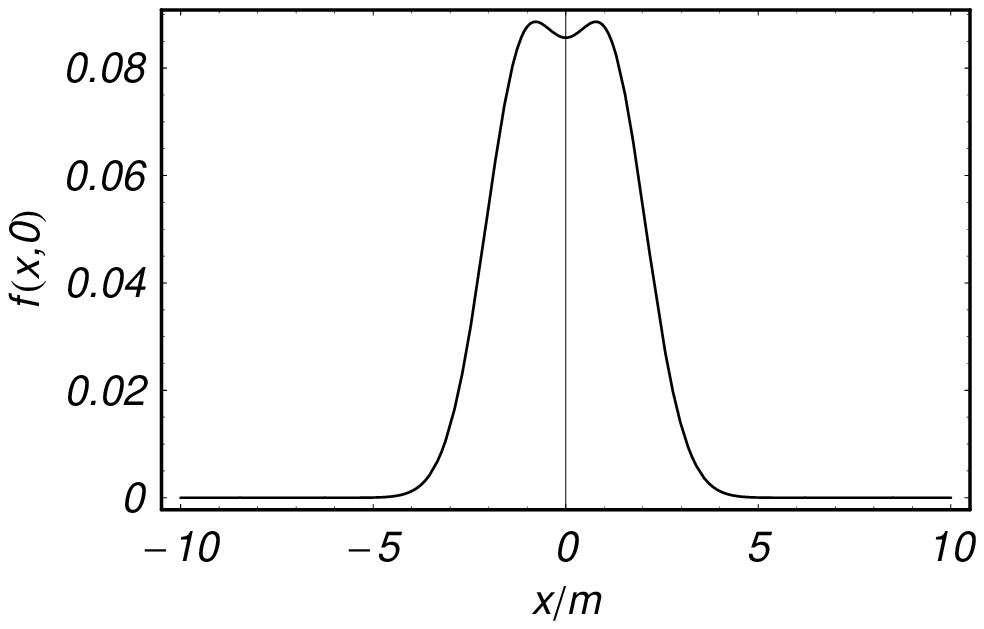}}
\subfigure[$\theta=75^\circ$]{\label{fig:f75}
\includegraphics[width=0.5\textwidth]{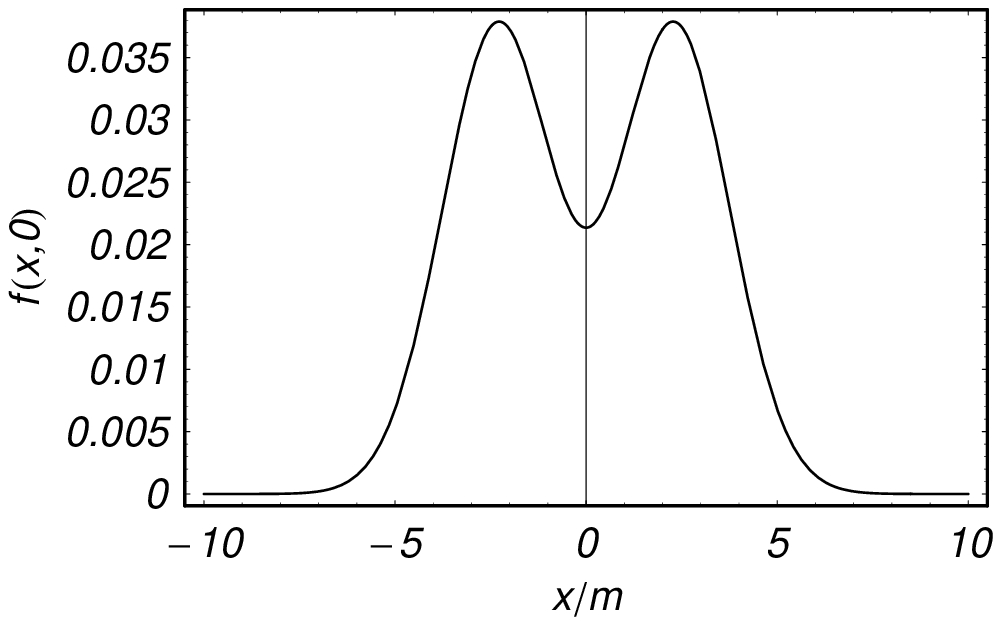}}
\subfigure[$\theta=80^\circ$]{\label{fig:f80}
\includegraphics[width=0.5\textwidth]{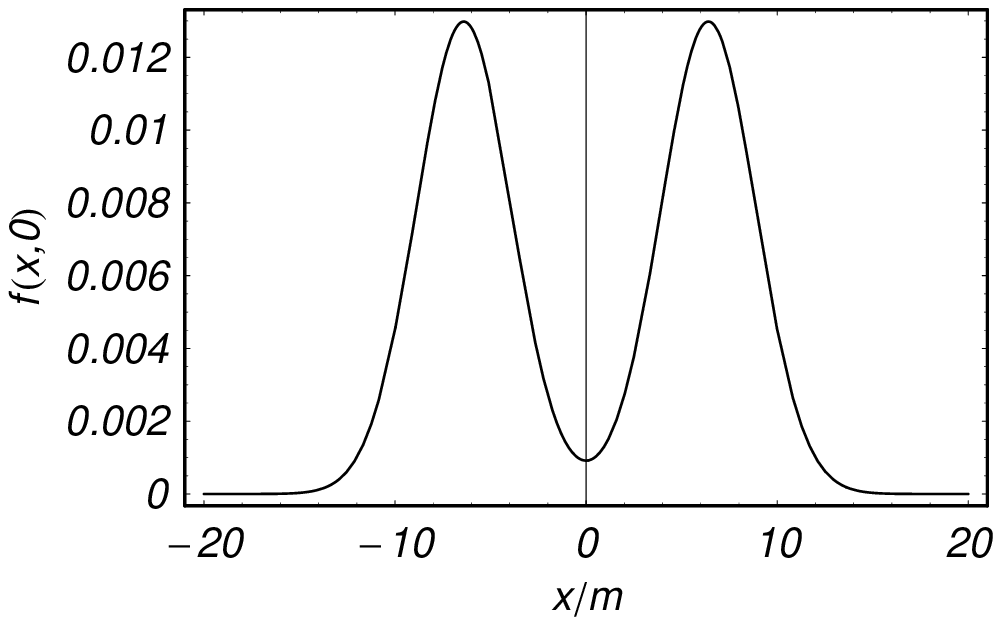}}
\subfigure[$\theta=85^\circ$]{\label{fig:f85}
\includegraphics[width=0.5\textwidth]{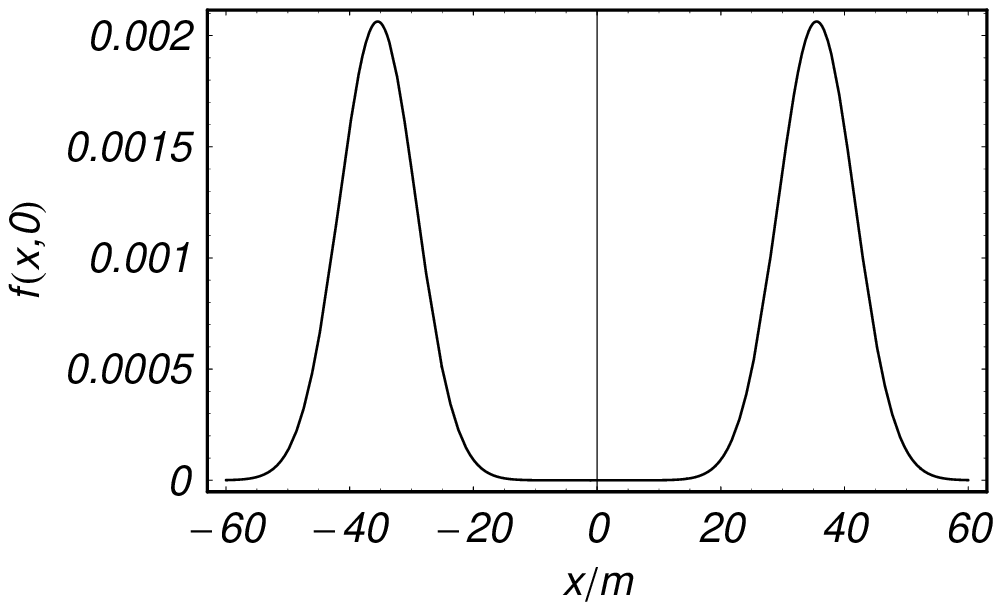}}
\caption{Lateral distribution of high energy muons in the transverse
plane as a function of $x$ coordinate with $y=0$, for fixed muon
energy around $10^{4.5}$\,GeV in air showers with primary energy
$E_0=10^{11}$\,GeV and at different zenith angles $\theta=
70^\circ,\ 75^\circ,\ 80^\circ,\ 85^\circ$ respectively.}
\label{fig:f}
\end{figure}

Fig.~\ref{fig:s70} shows a case with shower primary energy
$E_0=10^{11}$\,GeV and zenith angle $\theta=70^\circ$. It can be
seen that the geomagnetic deviation $s$ begins to dominate at high
energies, just as we predicted earlier. The transition point is
$s=2\sigma$, which can be read from the graph to be at an energy of
about $10^{3.5}$\,GeV. Fig.~\ref{fig:f70} shows the corresponding
lateral distribution of high energy muons in the transverse plane,
where positive and negative muons should form their own lobes
respectively. As our criterion predicts, their separation begins to
be distinguishable when muon energies are higher than
$10^{3.5}$\,GeV. Each type of muon with such energy would have
little chance ($<14\%$) of arriving and being found in the other
lobe, so that we can confidently distinguish the muon charges
provided that their relative positions in the transverse plane are
recorded with a good resolution.

Actually $\theta=70^\circ$ acts like a critical zenith angle, in
which case the curves of $2\sigma$ and $s$ nearly coincide for muon
energies $\gtrsim$ TeV. For larger zenith angles, the separation $s$
far exceeds $2\sigma$ in the whole energy range above TeV. Typical
examples are shown in Figs. \ref{fig:s75}, \ref{fig:s80} and
\ref{fig:s85} with zenith angles $\theta=75^\circ$, $80^\circ$ and
$85^\circ$ respectively. Accordingly, the two lobes of either
positive or negative muons can be easily recognized from each other,
as seen from Fig.~\ref{fig:f}.

Note that now the lateral distribution function Eq.~(\ref{eq:f})
depends on the specific muon energy, and will be denoted by
$f(x,y;E)$. From Eq.~(\ref{eq:s}), the centers of the two lobes
become very close to each other at highest energies. Consequently
there will be a technical upper limit to the muon energy, above
which the two lobes are hardly distinguishable due to the limited
detector resolution. Luckily, as Fig.~\ref{fig:N80} will show, the
number of muons produced at such highest energies is virtually zero,
hence this energy range will be excluded from our consideration of
muon production and deviation.

Note also that with the growth of zenith angle, the separation
between positive and negative muons becomes larger, so that the
demand for detector resolution is more accessible. Therefore, in
order to detect an unambiguous geomagnetic separation between the
positive and the negative high energy muons, we should focus our
attention on EAS events with zenith angles $> 70^\circ$. On the
other hand, for almost horizontal air showers with a nonzero
azimuth, the two lobes of muons would travel very different
distances to reach the ground, resulting in extra asymmetry in the
positive and negative muon distributions \cite{Ave:2000xs}.
Therefore, we suggest an optimum zenith range $75^\circ \leq \theta
\leq 85^\circ$ for our approach, which contributes over $17\%$ of
the half total solid angle.

\section{Muon Number Density}
\label{sec:num}

We note that the lateral distribution we calculated in the early
section is only for muons with fixed energies. To see the actual
muon number density in an extensive air shower, we need to know the
muon energy spectrum, of which we can get a rough estimate through
the same Heitler model.

We start by considering the number of charged pion produced in the
development of an air shower. From Eq.~(\ref{eq:En}) we have
\begin{equation}
N_n = \frac{E_0}{E_n} \left( \frac{2}{3} \right) ^n, \label{eq:Nn}
\end{equation}
or, eliminating $n$ by Eq.~(\ref{eq:En1}) and adopting best-fit
values of the parameters,
\begin{equation}
N(E) = \frac{E_0}{E} \left( \frac{2.870 + \lg E}{2.870 + \lg E_0}
\right) ^{1.656}. \label{eq:NE}
\end{equation}
In this expression we have assumed the proliferation of charged
pions to be a continuous process, during which the number of charged
pions grows while their energy drops simultaneously. We can
calculate the number of pions produced in the energy range $E$ to
$E-dE$ by taking the derivative of $N(E)$,
\begin{equation}
dN = \left( -\frac{dN(E)}{dE} \right) dE. \label{eq:dN}
\end{equation}

\begin{figure}
\begin{center}
\includegraphics{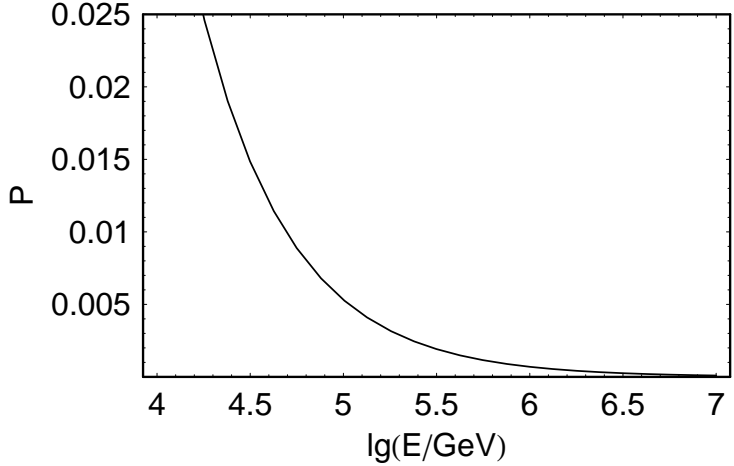}
\caption{Estimated probability of pion decay at high energies,
calculated for air showers with primary energy $E_0=10^{11}$\,GeV
and zenith angle $\theta=80^\circ$.} \label{fig:P80}
\end{center}
\end{figure}

As mentioned in section 2, the parent pions of these newly produced
ones actually have a small chance of decaying to muons instead of
interacting. We can estimate this probability by taking the ratio of
the interaction length to the decay mean free path,
\begin{equation}
P(E) \simeq \frac{\lambda_I(E)}{c\gamma \tau_{\pi^{\pm}} \rho(H)} =
\frac{h_0 m_{\pi^{\pm}}}{c \tau_{\pi^{\pm}} E}
~\frac{\lambda_I(E)}{X(E) \cos\theta} = \frac{115}{E}
~\frac{145-10.5 \lg E}{X(E) \cos\theta}, \label{eq:PE}
\end{equation}
where we have used the relation between the atmospheric depth $X(H)$
and the isothermal atmospheric density $\rho(H)$. Since the
probability of pion decay is very small with energies above
$10^4$\,GeV (see Fig.~\ref{fig:P80}),
we can safely neglect the charged pions lost by decaying in the
Heitler model before the energy reaches too low. Thereby the number
of high energy pions that decay in the energy range $E$ to $E-dE$ is
\begin{equation}
N_{dec}(E)dE = P(E) \, \frac{1}{N_{ch}(E)} \left( -\frac{dN(E)}{dE}
\right) dE, \label{eq:Ndec}
\end{equation}
where the multiplicity of charged particles $N_{ch}(E)$ is given by
Eq.~(\ref{eq:Nch}).

\begin{figure}
\begin{center}
\includegraphics{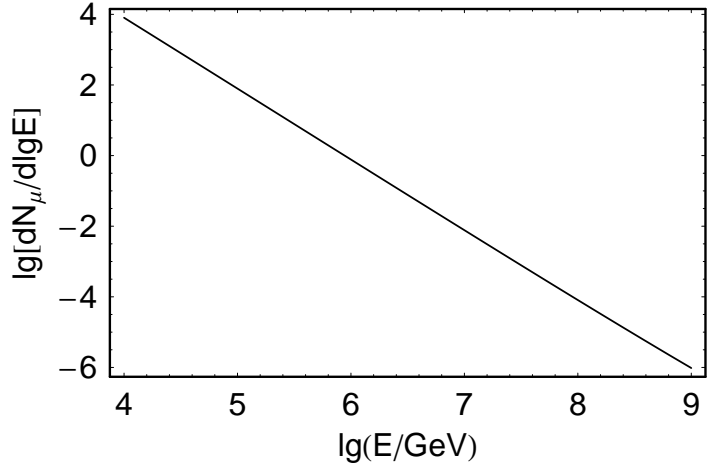}
\caption{The energy spectrum of high energy muons in air showers
with primary energy $E_0=10^{11}$\,GeV and zenith angle
$\theta=80^\circ$, expected by the revised Heitler model.}
\label{fig:D80}
\end{center}
\end{figure}

The muons from these pion decays typically lose several tens of
GeV's energy to ionization before reaching the ground, which is
negligible compared to their kinetic energies well above TeV
\cite{Knapp:2002vs}. Hence their energy spectrum $D_\mu(E)$ remains
virtually the same before they reach the ground. Allowing for an
energy fraction factor in pion decay, we have
\begin{equation}
D_\mu(E) = 1.27\, N_{dec}(1.27\,E). \label{eq:Dmu}
\end{equation}
Fig.~\ref{fig:D80} shows the calculated result for $D_\mu(E)$
with shower primary energy $E_0=10^{11}$\,GeV and zenith angle
$\theta=80^\circ$, exhibiting a perfect power-law $D_\mu \propto
E^{\gamma}$ with index $\gamma \approx -3$.

\begin{figure}
\begin{center}
\includegraphics{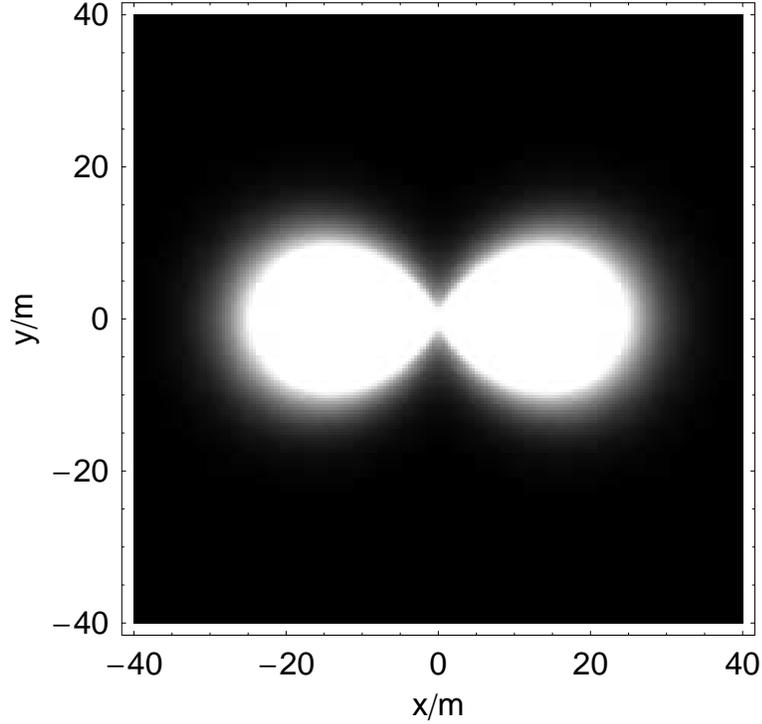}
\caption{Contour plot: number density of high energy muons in the
transverse plane for air showers with primary energy
$E_0=10^{11}$\,GeV and zenith angle $\theta=80^\circ$, calculated
with the revised Heitler model.} \label{fig:Dens80}
\end{center}
\end{figure}

Let us integrate Eq.~(\ref{eq:Dmu}) to see the number of muons with
energies $\ge E$,
\begin{equation}
N_\mu(E) = \int_E^{E_0} D_\mu(E) dE, \label{eq:Nmu}
\end{equation}
where $E_0$ is the energy of the primary cosmic ray particle.
\begin{figure}
\begin{center}
\includegraphics{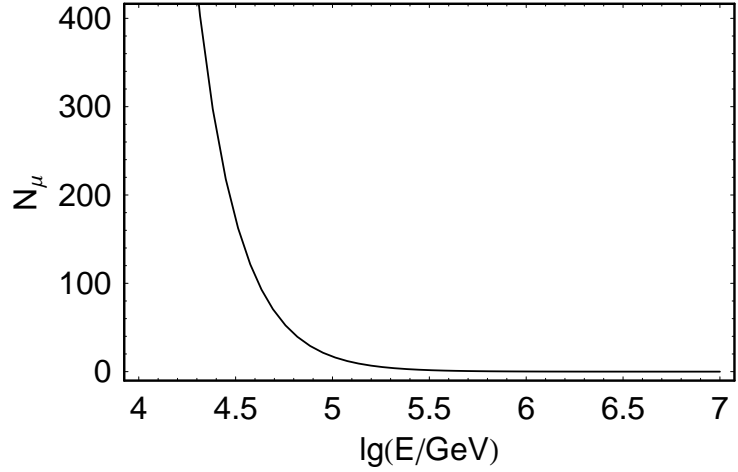}
\caption{Expected number of high energy muons in the revised Heitler
model for air showers with $E_0=10^{11}$\,GeV and
$\theta=80^\circ$.} \label{fig:N80}
\end{center}
\end{figure}
Moreover, with the muon energy spectrum, we can calculate the
``integral'' number density of high energy muons in the transverse
plane. Referring to Eq.~(\ref{eq:f}) for the lateral distribution of
muons with energy $E$, let us multiply it by the spectrum $D_\mu(E)$
and integrate,
\begin{equation}
g(x,y) = \int_{E_b}^{E_1} f(x,y;E) D_\mu(E) dE, \label{eq:g}
\end{equation}
where we have chosen the lower bound of integration to be $E_b =
10^4$\,GeV, and the upper bound $E_1 \approx 10^{5.6}$\,GeV is given
by $N_\mu(E_1)=1$, above which energy there is no muon produced. The
numerically calculated muon number density $g(x,y)$ for
$\theta=80^\circ$ is shown in Fig.~\ref{fig:Dens80}, which can be
compared with simulation or experiment results.

As expected, the number of high energy muons is very limited
(Fig.~\ref{fig:N80}). So the detector array will actually record
several high energy muons arriving at discrete lateral positions,
instead of a continuous density distribution in the transverse
plane. Nevertheless, once we have determined the shower core
position, our calculated double-lobed distribution
(Fig.~\ref{fig:Dens80}) predicts that the high energy muons arriving
at each side of the shower core most probably ($>99\%$) belong to
the lobe on the same side, carrying the corresponding charge. The
shower core can be located by analyzing the density distribution and
energy deposits of plenty lower energy secondary particles
\cite{Apel:2005ch}. Since we know which side for which type of muon
charge from the direction of local geomagnetic field, in this way we
can confidently identify these high energy muons with their own
charges.

Our final purpose is to obtain the charge ratio $R_\mu$ for such
high energy muons, which will give us valuable information on the
composition of primary cosmic ray particle and the understanding of
hadro-production processes. However, we need to point out here that
it is impractical to measure the charge ratio on an event-by-event
basis. Because of the stochastic nature of extensive air showers,
the fluctuation in muon numbers will be large compared to the
limited total number of high energy muons in a single event. Also,
not all the high energy muons in a particular event will be
detected, since the array detectors do not cover the whole region.
Therefore, both positive and negative muon numbers should be
collected from a large number of inclined EAS events, so that we can
get a statistically sound value of $R_\mu$ for high energy muons.

\section{Summary}
\label{sec:sum}

In this article we analyzed the possibility of obtaining the charge
information of high energy muons in very inclined extensive air
showers. We have demonstrated that positive and negative high energy
muons in sufficiently inclined air showers can be distinguished from
each other through their opposite geomagnetic deviations in the
transverse plane. We developed a revised Heitler model to calculate
this distinct double-lobed distribution, and studied the condition
for the two lobes of either positive or negative muons to be
separable with confidence. From our criterion of resolvability, we
concluded that a zenith angle $75^\circ \leq \theta \leq 85^\circ$
will be most suitable for our approach.

There are already some results from full air shower simulations that
take into account the geomagnetic effect on muon propagation
\cite{Ave:2000xs}~\cite{Hansen:2004kf}. They illustrated remarkable
double-lobed muon lateral density profile in very inclined air
showers, which is in agreement with our expectation qualitatively.
However, no present study has fully considered the high energy part
of muon content, which can be used to compare with our results. Thus
we would like to propose future simulations of very inclined
extensive air showers that focus on the behavior of high energy
muons. They also have to keep track of the muon charges and the
relation to their lateral positions.

For our method to be applicable in measuring the high energy muon
charges, there are some requirements on the shower detector
performance. First, the muon detectors should be able to measure
muon energies up to over $10^4 - 10^5$\,GeV, so that we can sort out
those muons with energies high enough for our purpose. Moreover,
since the high energy muons arrive very near the shower core, we
would have to look for EAS events whose shower cores are inside the
coverage of the detector array. Besides, to distinguish the muons'
charges from their lateral positions, a detector array is expected
to have a high resolution in the shower transverse plane.

As far as we are concerned, there are still some technical
limitations to the above requirements. For example, the detectors
near the shower core are usually saturated with signals from lower
energy muons \cite{Anchordoqui:2004xb}. Therefore, we suggest to
employ muon detectors that are especially aimed at detecting high
energy muons in the shower array. They should have a high threshold
energy up to $10^4$\,GeV, so as to avoid unwanted signals from low
energy muons. Or these detectors may be extensions of existing ones
like water \v Cerenkov detectors, but with special triggers to sort
out high energy signals.

Our method not only works for air showers with primary energies over
the ``ankle'' ($\sim 10^{10}$\,GeV), it can also be applied to study
cosmic rays in the ``knee'' region. In such cases, the primary
energy of the cosmic ray particle is around $10^{6}$\,GeV, several
orders lower than the examples we studied above. Accordingly, the
energies of muons produced by early generation pions range from
$10^{2}$\,GeV to $10^{3}$\,GeV. Produced higher in the atmosphere
with lower energies, these muons experience larger lateral
deviations, and are more susceptible to the geomagnetic bending.
Thus we shall have an even better condition to use our method to
obtain the muon charge information. What is more, for such energies
below $1000$\,GeV, there are alternative ways to distinguish the
muon charges \cite{Rastin:1984nv,Vulpescu:1998hm}. By comparing the
results, these experiments can serve as a test to validate or rule
out the possible application of our approach.

\begin{ack}
This work is supported by Hui-Chun Chin and Tsung-Dao Lee Chinese
Undergraduate Research Endowment (Chun-Tsung Endowment) at Peking
University. It is also partially supported by National Natural
Science Foundation of China (Nos.~10421503, 10575003, 10528510), by
the Key Grant Project of Chinese Ministry of Education (No.~305001),
and by the Research Fund for the Doctoral Program of Higher
Education (China).

\end{ack}

\end{document}